\documentclass{ws-ijmpa}

\def\shat{\hat{s}}

\def\sp{{\mathbf S}_t}

\newcommand{\kh}{{{\bf \hat k}}}

\newcommand{\be}{\begin{equation}}
\newcommand{\ee}{\end{equation}}

\newcommand{\bea}{\begin{eqnarray}}
\newcommand{\eea}{\end{eqnarray}}
\newcommand{\bfig}{\begin{figure}}
\newcommand{\efig}{\end{figure}}
\newcommand{\bc}{\begin{center}}
\newcommand{\ec}{\end{center}}

\begin{document}

\markboth{W. Bernreuther, M. F\"ucker, Z. G. Si}
{Mixed QCD  and Weak Corrections to
$t\bar{t}$ Production by Quark-antiquark Annihilation
}

%
\catchline{}{}{}{}{}
%

\title{Mixed QCD and Weak Corrections to
$t\bar{t}$ Production by Quark-Antiquark Annihilation}

\author{\footnotesize 
\bf W. Bernreuther\,$^{a,}$,
M. F\"ucker \,$^{a,}$,
Z. G. Si\,$^{b,}$ }
\address{
$^a$Institut f\"ur Theoretische Physik, RWTH Aachen, 52056 Aachen, Germany\\
$^b$Department of Physics, Shandong University, Jinan, Shandong
250100, China}

\maketitle


\begin{abstract}
We report on our computation of the
mixed QCD and weak  corrections to $q {\bar q} \to t\bar{t}$ 
including $t, \bar t$ spin effects, 
in particular on single top spin asymmetries.
\end{abstract}

Top quark physics will 
play an important role in testing the Standard Model (SM) and searching for new
physics beyond the SM, once large $t\bar{t}$ data samples are available at the Tevatron and
at the LHC in the future. For this aim, the SM predictions should be 
known as precisely as possible. Specifically weak interaction contributions to
$t \bar t$ production should be taken into account. Although they are
nominally subdominant with respect to the QCD contributions they can
become important at large $t \bar t$ invariant mass due to large
Sudakov logarithms (for reviews and references, see e.g. 
\cite{Melles:2001ye,Denner:2001mn}).

SM weak interaction effects in hadronic production of heavy quark pairs
were considered previously.  
The parity-even and parity-odd order $\alpha_s^2
\alpha$ vertex corrections
were determined 
in \cite{Beenakker:1993yr} and in
\cite{Kao:1999kj}, respectively (see also \cite{Kao:1997bs}). 
In ref. \cite{Kao:1999kj}
also parity-violating non-SM effects were analysed.
The box contributions to $q {\bar q} \to t \bar t$ were not taken into
account in these papers. 
In the following we report on the results of 
the mixed QCD and weak radiative corrections of order $\alpha_s^2
\alpha$ to the  (differential) cross section
of  $t\bar t$ production by $q\bar q$ annihilation, keeping the full
information on the spin state of the $t\bar{t}$ system. 
These are necessary ingredients for definite SM predictions, 
in particular of parity-violating observables associated with the
spin of the (anti)top quark. 

Top quark pair production both at the Tevatron and at the LHC is
dominated by the QCD contributions to $q \bar q \to t \bar t$ and
$g g \to  t \bar t$. Their cross section including the full dependence
 on the $t, \bar t$ spins are known to order 
$(\alpha_s^3)$ \cite{Bernreuther:2001rq,Bernreuther:2004jv}. 
Due to color conservation,
the leading contributions involving
electroweak interactions are the order  $\alpha^2$  Born terms  for
$q \bar q \to t \bar t$ and the
mixed contributions of order  $\alpha_s^2 \alpha$.
For  the 
$q+\bar{q}\rightarrow t+\bar{t}$, and
$q + {\bar q} \rightarrow t+\bar{t} + g$ processes,
the inclusive spin-summed cross sections 
may be written,
to NLO in the SM couplings, in the form 
\begin{equation}
{\sigma}_{q \bar q} \; = \; {\sigma}_{q \bar q}^{(0)QCD} + 
\delta {\sigma}_{q \bar q}^{QCD} + \delta {\sigma}_{q \bar q}^{W} \; ,
\label{mixed}
\end{equation}
where the first and  second term are  the LO (order $\alpha_s^2)$ and
NLO (order $\alpha_s^3)$ QCD contributions 
\cite{Nason:1987xz,Beenakker:1990ma,Bernreuther:2000yn}. The third term is
generated by the electroweak contributions, and can be
decomposed as follows:
\begin{equation}
\delta {\sigma}_{q \bar q}^W(\shat,m^2_t) \; = \; \frac{4\pi \alpha}{m^2_t}[
\alpha \, f^{(0)}_{q \bar q}(\eta) + \alpha_s^2 \, f^{(1)}_{q \bar q}(\eta)]
\, ,~~~\eta = \frac{\shat}{4m^2_t} -1\, ,
\label{eq:xsection}
\end{equation}
\vspace*{-0.9cm}
\begin{figure}[hptb]
\begin{center}
\epsfig{file=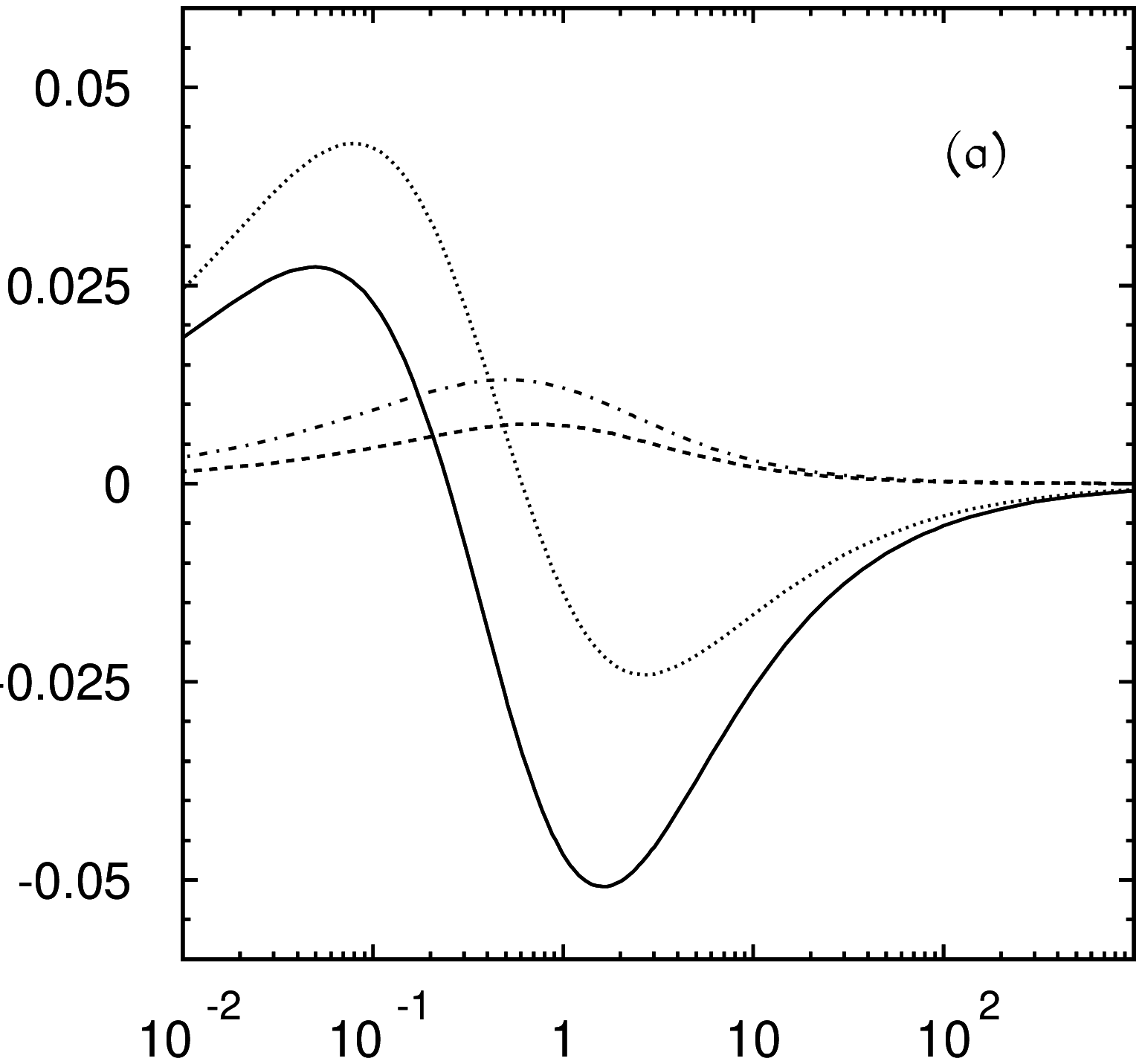, width=6cm ,height=6cm}
\epsfig{file=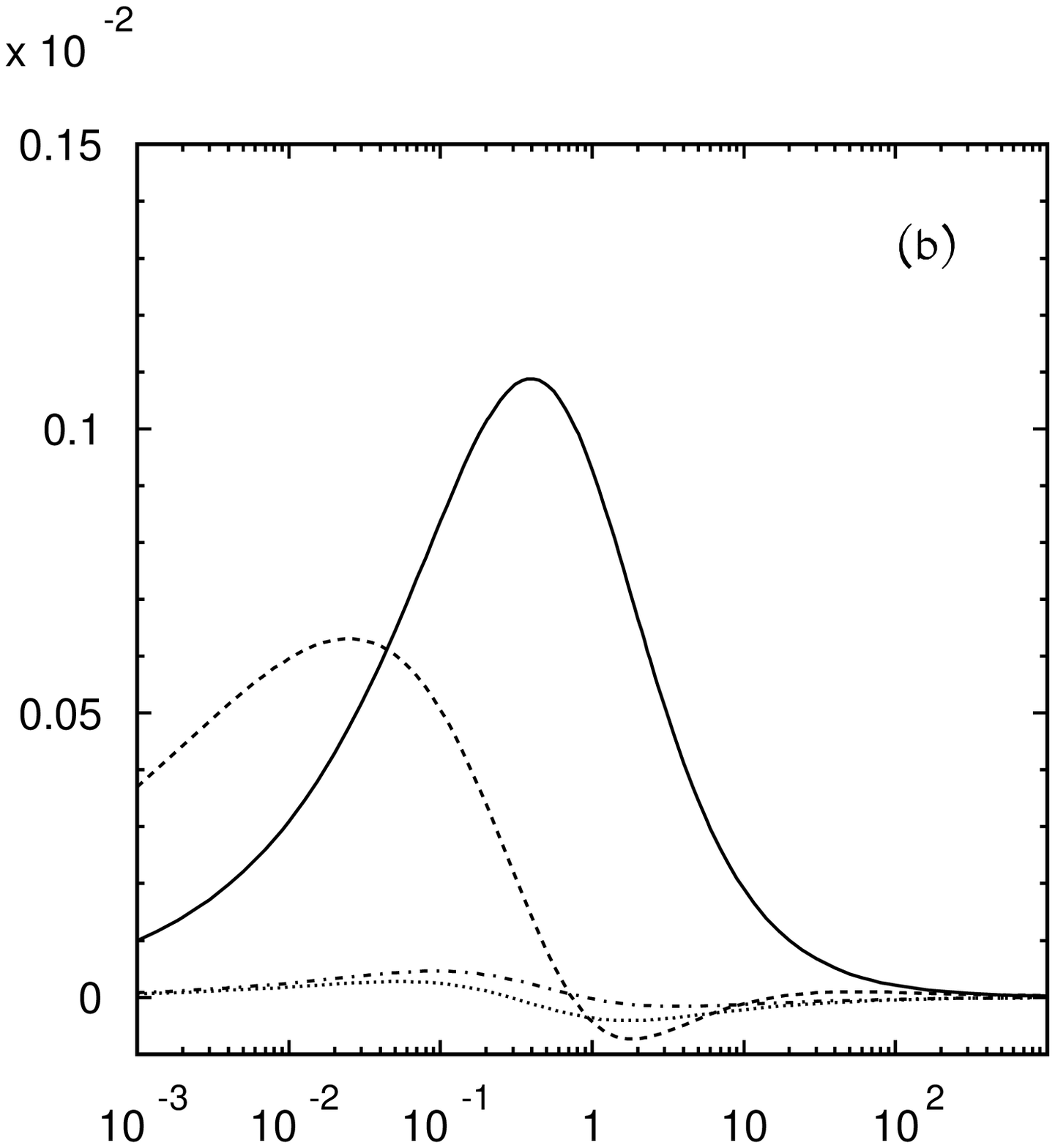, width=6cm, height=6cm}
\end{center} \vspace*{-0.5cm}
\caption{(a) Dimensionless scaling functions $f^{(0)}_{q \bar q}(\eta)$
(dashed), $f^{(1)}_{q \bar q}(\eta)$ (solid)  that determine
the parton cross section  (\ref{eq:xsection}) for
$q=d$ type quarks. The dash-dotted and dotted lines correspond to
the respective functions for $q=u$ type quarks. 
(b)  Contributions of the LO (solid) and NLO QCD (dashed)
contributions (taken from $^7$) 
 and of the 
mixed $\alpha_s^2 \alpha$ contributions (dotted and dash-dotted
line refers to
initial d-type and u-type quarks, respectively) to the cross section
(\ref{mixed}) 
in units of $1/m_t^2$.
}\label{fig:sigqq}
\end{figure}
where  ${\shat}$ is the $q \bar q$
c.m. energy squared.
In our calculations \cite{BFS05}
we used $m_t = 178$ GeV, $m_Z=91.188$ GeV,  
$sin^2\theta_W =0.231$, and 
 $m_H =114$ GeV for the mass of the SM Higgs boson. 
In Fig. \ref{fig:sigqq}(a) the  functions $f^{(i)}_{q \bar q}$ are
displayed 
as functions of $\eta$ for 
annihilation of initial massless partons $q \bar q$ of the first and
second generation with weak isospin $\pm 1/2$. As expected the 
$\alpha_s^2 \alpha$ corrections are significantly larger than the the lowest
order photon and $Z$ boson exchange contributions. 
The correction (\ref{eq:xsection}) to the
$q \bar q$ cross section  has recently been computed also 
by \cite{KSU}. We have compared our results and find excellent
numerical agreement. 
The contribution of the  corrections shown in 
Fig. \ref{fig:sigqq}(a)  to the $t \bar t$ cross section at the Tevatron is very small.
This is mainly due to the fact that the 
order $\alpha_s^2 \alpha$ corrections change sign for larger $\eta$.
The significance of the contributions can be enhanced by suitable
cuts, e.g. in the $t {\bar t}$ invariant mass.
In Fig. \ref{fig:sigqq}(b) the order $\alpha_s^2$, $\alpha_s^3$,  and
the $\alpha_s^2 \alpha$ contributions to the cross section
(\ref{mixed}) are shown as functions of $\eta$.  In these plots
$\alpha_s(m_t)=0.095$ and $\alpha(m_Z) = 0.008$ was chosen.
One sees that for $\eta \gtrsim 1$
the mixed corrections become of the same size or 
larger in magnitude than 
the NLO QCD contributions, and at $\eta \sim 10$ the
$\alpha_s^2 \alpha$ contributions are already about 15 percent of the 
LO QCD cross section.
These regions
can be investigated by studying the  distribution of the
$t {\bar t}$ invariant mass $M_{t \bar t}$. 

Next we consider spin asymmetries. Denoting the top spin operator by
$\sp$ and its projection onto an arbitrary unit axis ${\bf\hat a}$
by $\sp \cdot {\bf\hat a}$ we can express 
its unnormalized  partonic expectation
value, which we denote by double brackets, in terms of the
difference between the ``spin up'' and ``spin down'' cross sections:
\begin{equation}
  2 \langle  \langle \sp \cdot {\bf\hat a}  \rangle  {\rangle}_i
  = \sigma_i(\uparrow ) - \sigma_i(\downarrow) \,  .
\label{sspas}
\end{equation}
Here $i = q \bar q$ and the arrows refer to the spin state of the top quark 
with  respect to ${\bf{\hat a}}$. An analogous formula  holds for the antitop
quark. It is these expressions  that enter the 
predictions for (anti)proton collisions.

There are two types
of single spin asymmetries (\ref{sspas}):
parity-even, T-odd asymmetries and
parity-violating, T-even  ones. The asymmetry associated with
the projection $\sp$ onto the normal of the $q, t$ scattering
plane belongs to the first class.
It is induced by the absorptive part of the amplitude which receives
a QCD contribution at NLO \cite{Bernreuther:1995cx,Dharmaratna:xd}. 
The weak
contribution is even smaller  than the one from QCD which is of the
order of a few percent. For the sake of brevity we do not display it here.
\vspace*{-1.1cm}
\begin{figure}[hptb]
\begin{center}
\epsfig{file=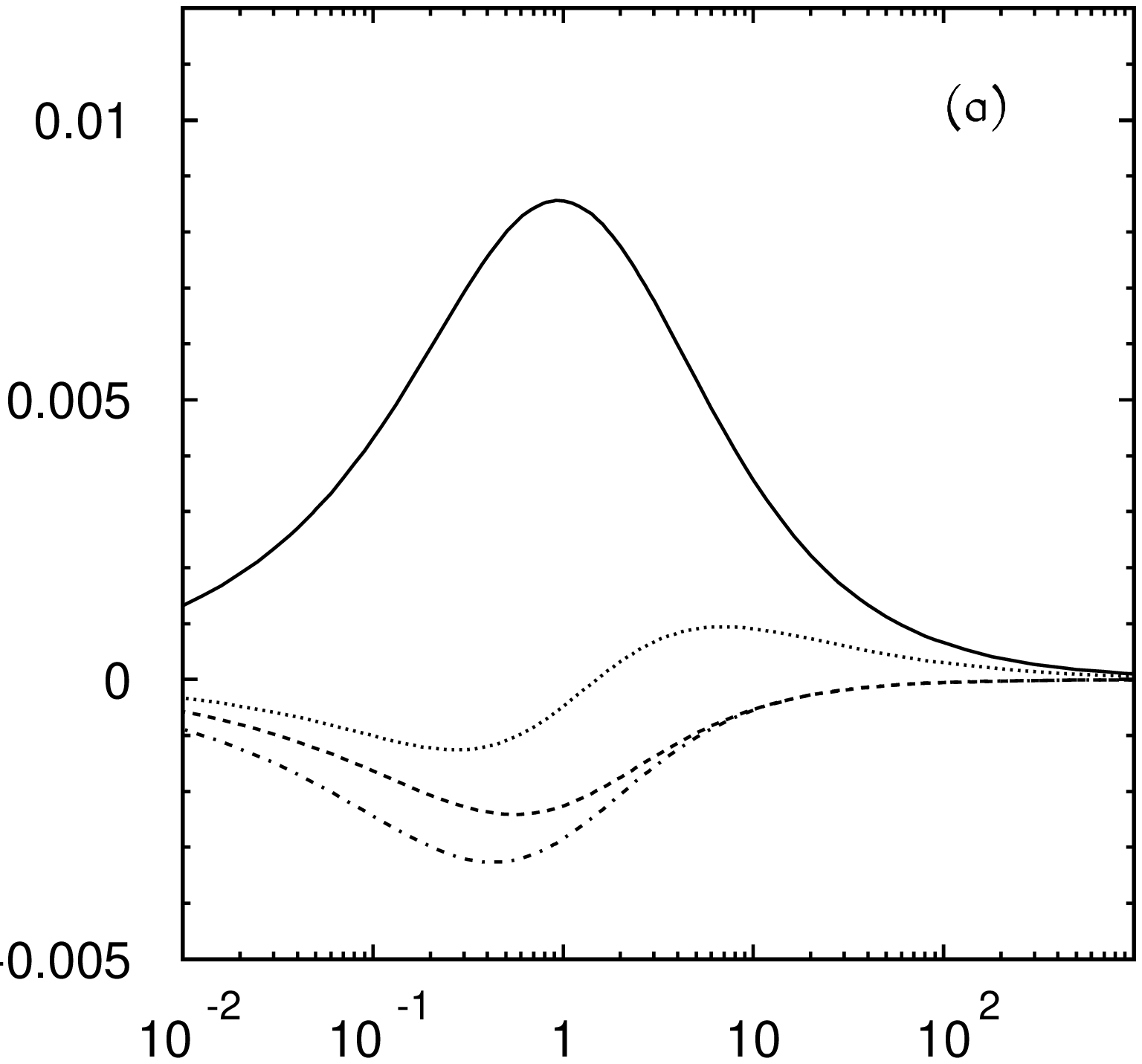, width=6cm,height=6cm }
\epsfig{file=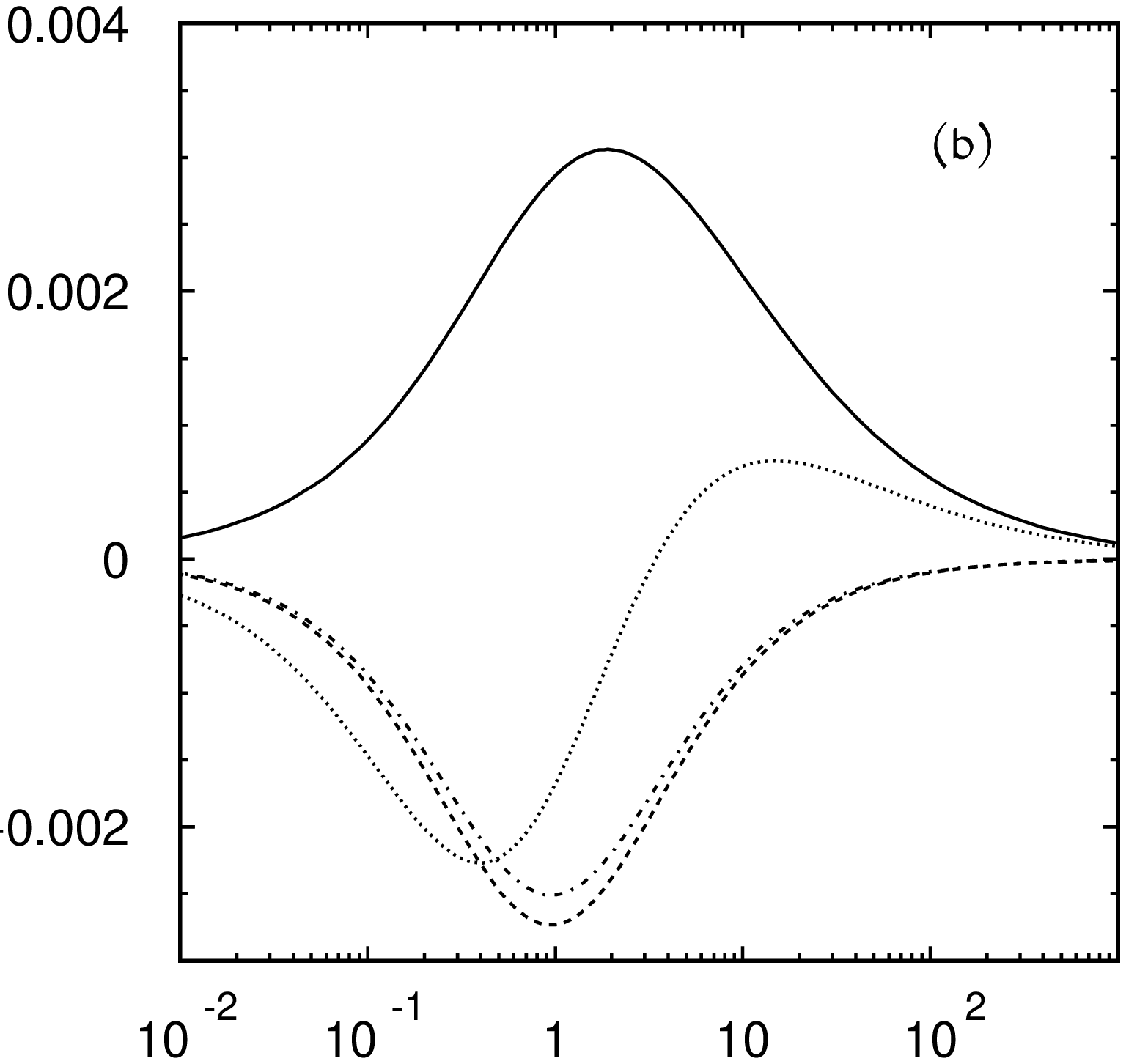, width=6cm,height=6cm }
\end{center}
\vspace*{-0.6cm}
\caption{Scaling functions $h^{(0,a)}_{q \bar q}(\eta)$
(dashed), $h^{(1,a)}_{q \bar q}(\eta)$ (solid)  that determine
the expectation value   (\ref{eq:sinspin}) for (a) beam  and (b) helicity  axis
 in the case
of $q=d$ type quarks. The dash-dotted and dotted lines correspond to
the respective functions for $q=u$ type quarks. }\label{fig:qqbeam}
\end{figure}
The P-odd, T-even  single spin asymmetries correspond to projections of
the top spin onto a polar vector, in particular onto 
an axis that lies in the scattering plane. Needless to say, they
cannot be generated within QCD; the leading SM contributions to
these asymmetries are the parity-violating pieces. 
We consider top spin projections onto the beam axis (which is relevant
for the Tevatron), and onto the helicity axis (of relevance for the LHC).
Naively, one might
define these  axes in the c.m. frame of the initial partons. However,
the observables  $\sp \cdot {\bf\hat a}$ are then not
collinear-safe. (The problem shows up once second-order QCD
corrections are taken into account.) A convenient frame with respect
to which collinear-safe spin projections can be defined is the $t \bar
t$ zero-momentum-frame (ZMF) \cite{Bernreuther:2004jv}.
With respect to this frame we define the axes  as 
${\bf\hat a} = {\bf\hat p}$ (beam basis) and ${\bf\hat a} = \kh$ (helicity axes)
where  ${\bf\hat p}$ is the direction of flight of one of the colliding
hadrons, and $\kh $ denotes the direction of
flight of the top quark.
The unnormalized  expectation values of 
${\bf S_t}\cdot {\bf\hat a}$ are again conveniently expressed by scaling
functions
\begin{equation}
\langle \langle 2 {\sp}\cdot {\bf\hat a} \rangle
{\rangle}_{q \bar q}  =\frac{4\pi \alpha}{m^2_t}[
\alpha \, h^{(0,a)}_{q \bar q}(\eta) + \alpha_s^2 \, h^{(1,a)}_{q \bar q}(\eta)]
\, .
\label{eq:sinspin}
\end{equation}
The results for the scaling functions corresponding to beam and helicity
axes are shown in Figs. \ref{fig:qqbeam}(a) and (b).
For the beam axes, and for
weak isospin $I_W =-1/2$ quarks the
$\alpha_s^2 \alpha$ corrections are significantly larger than the
LO values. This is in contrast to the helicity basis where the LO and
NLO terms shown in Fig. \ref{fig:qqbeam}(b) are of the same order of
magnitude. Moreover, in this basis the LO and NLO terms cancel each
other to a large extent for $I_W =-1/2$ quarks in the initial state.
Because $t \bar t$ production at the Tevatron occurs predominantly
by $q \bar q$ annihilation, these results imply that, when it comes to
searching for SM-induced parity-violating $t$ or $\bar t$ spin
effects, the reference axes of choice should be ${\bf\hat p}$.
The SM effect is quite small; for suitably chosen $M_{t \bar t}$ mass
bins one gets asymmetries of the order of 2  percent. This leaves a
large margin in the search for new physics contributions. 

To summarize, we have investigated  the
mixed QCD and weak interaction contributions to
the cross section of $q {\bar q}\to t {\bar t}$ and computed the
 parity-violating 
single (anti)top quark 
spin effects induced by these interactions
(for details, see \cite{BFS05}). The corresponding investigation 
of  $ gg \to t {\bar t}$, which is necessary for a comprehensive analysis
of $pp, \, p {\bar p} \to  t {\bar t}\, X$, is in progress.
The parity-violating observables associated with the
spin of the (anti)top quark are very sensitive
to non-SM parity-violating top quark interactions, 
and therefore they will become very important 
analysis tools  once sufficiently large $t
 \bar t$ data samples are available at the Tevatron and at the LHC.


\subsubsection*{Acknowledgements}
 This work was supported
by DFG SFB/TR9, by
DFG-Graduiertenkolleg RWTH Aachen, and by
NSFC and NCET of China.

\end{document}